%% file: edelhoff_DileptonTauCMS_HCP11.tex
\documentclass[epjCONF,columns]{svjour} 
\usepackage{graphics}
\usepackage[varg]{txfonts} 
\usepackage[latin1]{inputenc}
\usepackage[amssymb,thinspace]{SIunits}
\input{ownCommands.tex}
\session-title{HadronColliderPhysics Symposium 2011}
\begin{document}
\title{Search for supersymmetry in events with two leptons including a tau} \author{Matthias
  Edelhoff\inst{1} \fnmsep\thanks{\email{edelhoff@cern.ch}} (on behalf of the CMS Collaboration) }
\institute{${}^{1}$ I. Physikalisches Institut B,
RWTH Aachen,
Sommerfeldstr. 14,
D-52074 Aachen}
\abstract{ Searches for new physics in events with hadronic jets, missing transverse energy, and two
  leptons of which at least one is a hadronically decaying tau are presented. The result is based on
  a data sample corresponding to an integrated luminosity of 1 $\fbi$ at a center-of-mass energy of
  7 TeV collected by the CMS experiment at the LHC. No significant excess with respect to the
  standard model predictions is found.  }
%
\maketitle
\section{Introduction}
\label{intro}
This article summarizes searches for physics beyond the Standard Model (BSM), analyzing $\int L dt
\approx 1 \fbi$ of data recorded by the Compact Muon Solenoid Experiment (CMS).  Proton-proton
collisions where provided by the LHC at center of mass energies of $\sqrt{s} = 7 \TeV$ in 2011 .

CMS conducted several searches for BSM in finale states characterized by large missing transverse
energy (\MET) and hadronic activity. On the one hand, the high \MET\ signature occurs in models with
weakly interacting particles that escape detection, which are favored by cosmological measurements.
On the other hand, hadronic activity accours naturally in colored particle
interactions dominating BSM cross sections in proton-proton collisions. In this article we focus on
finale states containing a combination of two leptons, at least one of which is required to be a
hadronically decaying tau (e\tauh, $\mu$\tauh, or \tauh\tauh). The charges of these two leptons can
be either of same sign \cite{SUS-11-010} or opposite sign \cite{SUS-11-007}, leading to different
selection and background estimation strategies. The study concerning same and opposite sign finale
are described in sections \ref{sec:SS} and \ref{sec:OS}, respectively.

In both cases hadronic jets and \MET are reconstructed using the particle flow technique
\cite{PFlow} and jets are clustered using the anti-kt algorithm \cite{ak5}. The amount of hadronic
activity in the event is measured by the quantity $\HT = \sum p_T^{jet}$ and the requirement of two
or more jets per event. Hadronically decaying tau leptons (\tauh) are identified using the HPS
algorithm \cite{TAU-11-001}. To suppress multi-jet QCD backgrounds all leptons are required to be
isolated.

%

\section{The CMS Detector}
\label{sec:cmsDetector}
The central feature of the Compact Muon Solenoid (CMS) apparatus is a superconducting solenoid, of
6~m internal diameter, providing a field of 3.8~T. Within the solenoids volume are the silicon pixel and
strip tracker, the crystal electromagnetic calorimeter (ECAL) and the brass/scintillator hadron
calorimeter (HCAL). Muons are measured in gas-ionization detectors embedded in the steel return
yoke. In addition to the barrel and endcap detectors, CMS has extensive forward calorimetry. A much
more detailed description of CMS can be found elsewhere~\cite{JINST}.

\section{Same Sign Search}
\label{sec:SS}
Finale states with two leptons of the same charge are rare in the Standard Model (SM). Thus, the main
backgrounds for this search are quark or gluon jets misidentified as a \tauh\ (e.g. in
W+jets events) and events where the charge of one of the leptons is misidentified (e.g. in
dileptonic \ttbar\ events). Both backgrounds are estimated directly from data as described in
sections \ref{sec:backgrounds_fake} and \ref{sec:backgrounds_charge}. The influence of light leptons
not produced in the hard scattering (e.g. from heavy flavor decays) is small compared to that of
misidentified \tauh\, due to the abundance of hadronic jets in the region of interest. Contributions
from rare same sign SM processes such as diboson production, double W strahlung, or double parton
scattering are small and estimated using simulation.

\subsection{Event Selection}
\label{sec:SSSelection}
The same sign search region is selected requioring $\HT > 350~\GeV$ and $\MET > 80~\GeV$. For the
requiroment of two or more jets and in calculating \HT\, jets with transverse momenta $p_T > 40$ GeV
are considered. Leptons are required to be in $|\eta| < 2.4$ and have transverse momenta of $p^{e}_T
> 10~GeV, p^{\mu}_T > 5~GeV, p^{\tauh}_T > 15~GeV$, to ensure efficient trigger selection.

\begin{table*}[t]
\caption{Predicted backgrounds and observed event yields in the search region ($\HT > 350~\GeV$ and
  $\MET > 80~\GeV$). Statistical and systematic errors have been added in quadrature. The upper
  limit is set using the \cls method.}
\label{tab:SSResults}       
\begin{center}
  \begin{tabular}{l|ccc|c|c}
    \hline\noalign{\smallskip}

    & $e\tauh$    & $\mu\tauh$ & $\tauh\tauh$      & Total  & 95\% CL \\
    &     &    &   &   & UL yield \\ \hline 
    {Predicted background } & {$1.1\pm0.4$} & {$1.8\pm1.4$} & {$0.0\pm 0.2$} & {$2.9\pm 1.7$}&  \\
    {Observed} & {1} & {2} & {0} & {3} & {5.8}\\

    \noalign{\smallskip}\hline
  \end{tabular}
\end{center}
\end{table*}

\subsection{Estimating Misidentified Jet Contribution}
\label{sec:backgrounds_fake}
The HPS \tauh\ identification algorithm distinguishes hadronic jets created in the decay of a $\tau$
lepton from those created in the hadronisation of a quark or gluon by means of isolation and
reconstructed particle content. Nonetheless, the selection of hadronically decaying $\tau$ leptons
always includes a remaining contamination by misidentified quark or gluon jets.

In order to estimate this contamination we employ the tight-to-loose method (TL). First, we define a
loose \tauh\ selection by relaxing the isolation requirement, in addition to the tight
\tauh\ selection used in the analysis. Second, we measure the fraction $f_{TL}$ of loose
candidates, which pass the tight criteria in a sample containing predominately quark and gluon
jets. Finally, we extrapolate the expected number of misidentified tight \tauh\ candidates from the
number of observed loose candidates in the signal region.

The tight-to-loose ratio $f_{TL}$ mainly depends on the transverse momentum and pseudorapidity of a
given \tauh\ candidate and is measured in bins of those variables. The difference in $H_T$ of the
region where $f_{TL}$ is measured and the search regions are minimised in the definition of the
loose selection.  The results of this procedure have been shown to be in good agreement with
background simulation.

\subsection{Estimating Misidentified Charge Contribution}
\label{sec:backgrounds_charge}
Backgrounds due to charge misidentification arise from the relative abundance of SM processes with
two leptons of opposite charge, at least one of which is an electron or hadronically decaying
$\tau$. The contribution of muons with misidentified charge is found to be negligible. The
misreconstruction of Electron charge occurs due to energy loss in the tracking volume. Furthermore,
$\tauh$ charge misidentification occurs in three prong \tauh\ decays, when a track form the
background is wrongly associated with the \tauh\ object.

To estimate the impact of these effects, we compare the number of opposite- and same sign dilepton
pairs near the Z resonance. The Drell-Yan (DY) signal is fitted in the dilepton invariant mass spectrum
alongside backgrounds form misidentified leptons and other SM processes. We identify the probability
$f_q^{\ell}$ of lepton charge misidentification as the ratio of dilepton pairs recontructed with the
same sign to those reconstructed with opposite sign.


For electrons we measure $f_q^e = 2 \cdot 10^{-4}$ ($3 \cdot 10^{-3}$) in the ECAL barrel
(endcap). Differences arrise due to differences in the amount of tracker material in front
of the ECAL crystals.

For three prong \tauh\ decays we measure $f_q^{\tauh} = 7.1 \pm 1.0_{stat.} \pm 2.5_{syst.} \%$.

Again results from this data driven background estimation are in agreement with background simulation.

\subsection{Results}
\label{sec:SSResults}
A summary of the predicted background and the observed yield in the search region is given in table
\ref{tab:SSResults}. We do not observe evidence of an event yield in excess of the SM based
predictions and set 95\% CL upper limits (UL) on the number of observed BSM events. The hybrid
frequentist-baysian \cls method \cite{CLS}  is applied, including nuisance parameters and the signal
strength maximizing the ratio of the signal with background and background only likelihoods.


\section{Opposite Sign Search}
\label{sec:OS}
In contrast to the same sign search there are several SM processes, such as DY and \ttbar\ decays,
with finale states containing two leptons of oposite charge. The background from DY processes can be
sufficiently suppressed by the choice of search region. However, contributions from \ttbar\ decays
remain. For the channels e\tauh\ and $\mu$\tauh\ those are estimated from data as described in section
\ref{sec:backgrounds_ptll}. Naturally backgrounds due to misidentified quark and gluon jets remain
and are estimated as described in section \ref{sec:backgrounds_fake}. Furthermore, the fully hadronic
finale state \tauh\tauh\ is treated differently than the other finale states: Here we define background
enriched sideband regions for each considered background. In this instance
$f^{\text{bkg}}_{TL}$ is taken from simulation. Backgrounds are predicted extrapolating
from the number of loose candidates in a background enriched sideband to the search region.

\subsection{Event Selection}
\label{sec:OSSelection}
\begin{figure}
  \begin{center}
    \resizebox{0.75\columnwidth}{!}{%
      \includegraphics{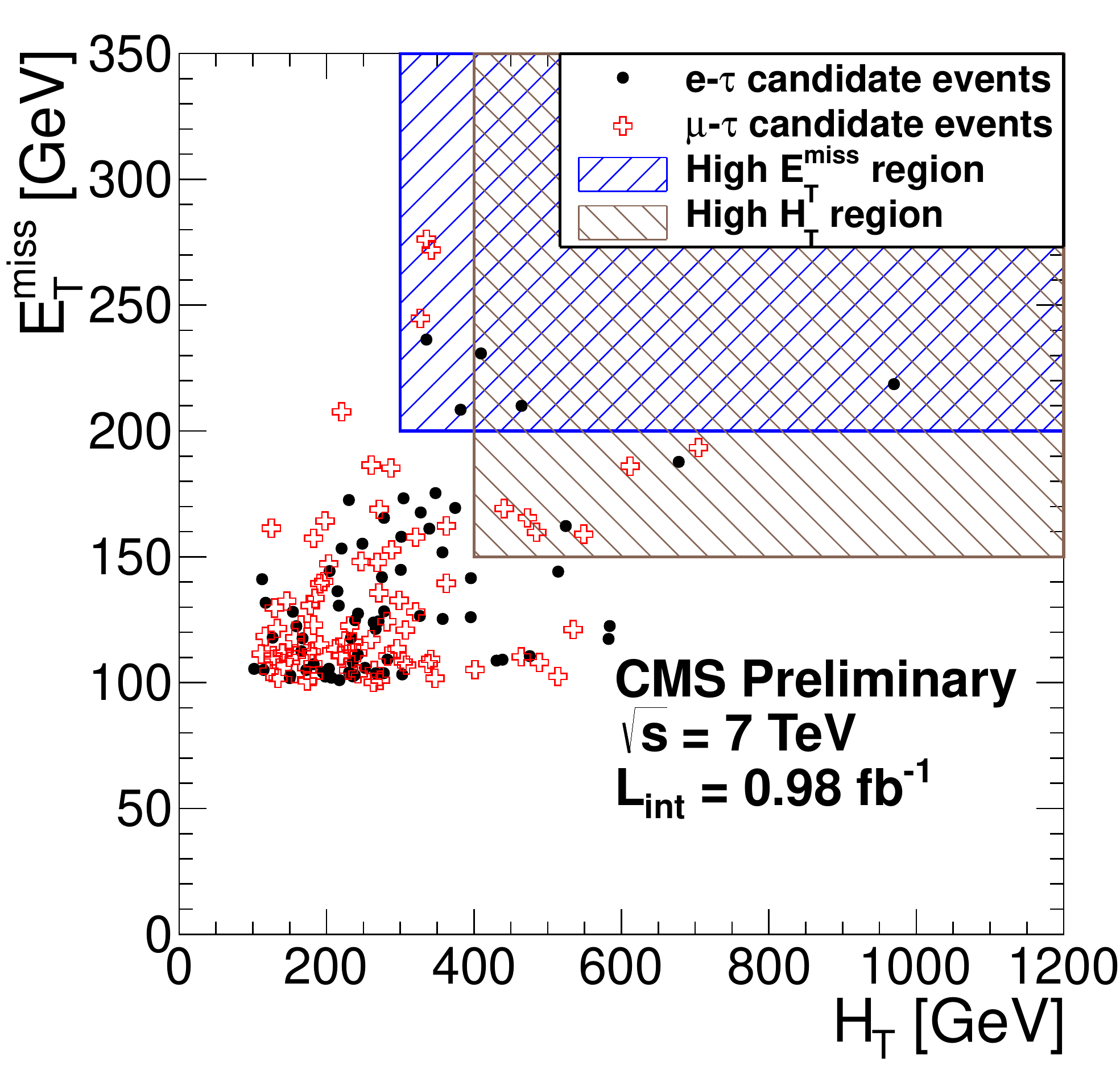} }
  \end{center}
  \caption{Distributions of \MET vs. \HT for data. The high \MET (high \HT) search region is
    indicated with the blue forward hatched (brown backward hatched) region.}
\label{fig:OShtmet}
\end{figure}

For the finale states containing a light lepton (e\tauh\ and $\mu$\tauh) two search regions are
defined (Fig. \ref{fig:OShtmet}). For brevity we focus on the high \MET\ region ($\HT > 300$ GeV
and $\MET > 200$ GeV). Jets with transverse momenta $p_T > 30$ GeV are considered in
$H_T$ and the two jet requiroment. All leptons are selected to have $|\eta| < 2.1$ and $p^{\ell}_T > 20$ GeV.

For the \tauh\tauh\ finale state two jets of transverse momenta $p_T > 100$ GeV are required and
instead of \MET\ the correlated variable $\MHT = |\sum \vec{p}^{\text{jet}}_T| > 200$ GeV is used,
for jets satisfying $p_T^{\text{jet}} > 30$ GeV. This is done to minimize turn-on effects due to the
trigger selection. Als the HPS \tauh\ identification and transverse momentum requirement
($p_T^{\tauh} > 15$ GeV) are relaxed, in order to maximize the statistical significance of the
result.

%
%

\begin{table}
\caption{Predicted backgrounds and observed event yields in the high \MET ($\HT > 300~\GeV$ and
  $\MET > 200~\GeV$) region for the opposite sign e\tauh and $\mu\tauh$ channels and for the \tauh\tauh\ search
  region ($\ge$ two jets with $p_T^{\text{jet}} > 100$ GeV and $\MHT > 200$ GeV). Errors are
  separated in statistical and then systematic uncertainty.}
\label{tab:OSResults}       
\centering
  \begin{tabular}{l|c|c}
    \hline\noalign{\smallskip}
& e$\tauh + \mu\tauh$ &  \tauh\tauh\\
\hline
\ptll\ Prediction & $5.9 \pm 1.5 \pm 1.9$ & ---\\
$\text{TL Prediction}$ & $1.6 \pm 0.6 \pm 0.2$ & ---\\
QCD                & --- & $0.58 \pm 0.02 \pm 0.41$ \\
W+Jets           & --- & $0.00 \pm 1.20 \pm 0.10$ \\
\ttbar   & --- & $2.18 \pm 2.18 \pm 0.35$ \\
$Z\to\nu\bar{\nu}+Jets$  & --- & $0.00 \pm 0.16 \pm 0.02$ \\ 
\hline
$\sum \text{Prediction from data}$ &  $7.5 \pm 1.6 \pm 1.9$ &  $2.76 \pm 2.50 \pm 0.55$\\
\hline
$\sum{SM}$ simulation & $9.6 \pm 4.2 \pm 3.1$ & $4.56 \pm 1.08 \pm 0.91$\\
\hline
$\text{Data}$ & $8$ & $3$\\
    \noalign{\smallskip}\hline
  \end{tabular}
\end{table}

\subsection{Estimating Dileptonic $\ttbar$ Contribution}
\label{sec:backgrounds_ptll}
We use the dilepton transverse momentum ($\ptll$) method to estimate the contribution of dileptonic
\ttbar\ events in the signal region of opposite sign e\tauh\ and $\mu$\tauh\ events. We
estimate the contribution of \ttbar\ events to the corresponding light lepton channels (ee, e$\mu$,
and $\mu\mu$), following the idea \cite{ptll} that the variable $\ptll$ can be used to model $\MET
= p_T(\nu\nu) $ \cite{SUS-11-011}. 

Here, we exploit the fact that in dileptonic \ttbar\ decays the $p_T$ distributions of the leptons are
related to those of the neutrinos via the common boosts from the intermediate top and W decays. This
relation is governed by the well understood W polarization, which can be reliably accounted for.

Contamination by events which stem from Z decays is first reduced by a $76 < m_{\ell \ell} < 106$
GeV and a $\MET > 50$ GeV requirement. The remaining contribution is then predicted and subtracted
using the same procedure as in Ref. \cite{top}. The bias of the \ptll\ distribution due to the \MET\
requirement is measured and accounted for.

Finally, lepton universality allows us to extrapolate from the light lepton channels to the \tauh\
channels in question. In this, \tauh\ reconstruction efficiency, acceptance, and branching ratios are
taken from simulation.


\subsection{Results}
\label{sec:OSResults}
A summary of the observed event yields and the data driven background predictions in the search
regions is given in table \ref{tab:OSResults}. We observe no excess of events over the SM
predictions. Also these predictions are shown to be in agreement with SM expectations from
simulation. 

We procede to evaluate three benchmark scenarios, referred to as LM1, LM2 and LM13, of the minimal
supersymmetric extension of the standard model (cMSSM) \cite{CMSSM}. The parameter values for [LM1,
LM2, LM13] are $m_0 = [60, 185, 270]$, $m_{1/2} = [250, 350, 218]$, $\tan \beta = [10,35,40]$, $A_0
= [0,0,-553]$, and $\mu > 0$ \cite{LMPoints}. We place 95\% confidence level upper limits on the
cross section of those scenarios using again the \cls\ method (Table \ref{tab:OSLimits}). The
combination of the three channels takes differences in search regions and correlations of the
uncertainties into account. All three scenarios are ruled out by the presented results.  Furthermore
we publish \cite{SUS-11-007} additional information to allow testing of specific BSM models against
these results.

\begin{table}
\caption{Summary of model cross sections as well as expected and measured $95\%$ upper limits as derived through the \cls method. }
\label{tab:OSLimits} 
\centering
  \begin{tabular}{l|c|cc}
    \hline\noalign{\smallskip}
Model &$\sigma_{\text{model}}^{\text{NLO}}$ [pb]  & UL $ \sigma_{expected}^{CL_S^{95\%}}$ [pb] & UL $ \sigma_{measured}^{CL_S^{95\%}}$ [pb]\\
\hline
LM1 & $6.6$ & $2.4 \pm 1.4$ & $2.8$\\
LM2 & $0.8$ & $0.6 \pm 0.3$ & $0.6$\\
LM13 & $9.8$ & $1.2 \pm 1.0$& $1.5$ \\
    \noalign{\smallskip}\hline
  \end{tabular}
\end{table}


\section{Conclusion}
\label{sec:conclusion}
Searches for physics beyond the standard model with \tauh\ final states using $\approx 1 \fbi$ of
integrated luminosity are summarized. Dominant backgrounds are estimated from the data taking the
challenges of the hadronic $\tau$ finale state into account. No deviation from the SM is found and
$95\%$ CL upper limits are computed. 

\end{document}

%% file: ownCommands.tex
\newcommand{\TeV}{\tera\electronvolt}
\newcommand{\GeV}{\giga\electronvolt}

\newcommand{\fb}{\femto\barn}
\newcommand{\fbi}{\ensuremath{\fb^{-1}}}

\newcommand{\MET}{\ensuremath{E^{miss}_T }}
\newcommand{\MHT}{\ensuremath{H^{miss}_T }}
\newcommand{\HT}{\ensuremath{H_T~}}
\newcommand{\ttbar}{\ensuremath{\text{t}\bar{\text{t}}}}

\newcommand{\ptll} {\ensuremath{p_T(\ell\ell)}}
\newcommand{\cls}  {\ensuremath{\mathrm{CL_S}}}
\newcommand{\tauh}  {\ensuremath{\tau_\text{h}}}